\documentclass[aps,prd,reprint,superscriptaddress]{revtex4-1}
\usepackage[utf8]{inputenc}

\usepackage{natbib,graphicx,amsmath,amssymb,bm,multirow,yfonts,color}
\usepackage[normalem]{ulem}
\newcommand{\be}{\begin{equation}}
\newcommand{\ee}{\end{equation}}
\newcommand{\bea}{\begin{eqnarray}}
\newcommand{\eea}{\end{eqnarray}}
\newcommand{\ba}{\begin{align}}
\newcommand{\ea}{\end{align}}

\renewcommand{\vec}[1]{\boldsymbol{#1}}

\newcommand{\gccm}{\,\mbox{g\,cm}^{-3}}
\newcommand{\mev}{\,\mbox{MeV}}

\def\lsim{\stackrel{\scriptstyle <}{\phantom{}_{\sim}}}
\def\gsim{\stackrel{\scriptstyle >}{\phantom{}_{\sim}}}

\begin{document}

\title{Impact of Nucleon-Nucleon Bremsstrahlung Rates Beyond One-Pion Exchange}

\author{A.\ Bartl}
\affiliation{Institut f\"ur Kernphysik, Technische Universit\"at Darmstadt, 64289 Darmstadt, Germany}
\affiliation{ExtreMe Matter Institute EMMI, GSI Helmholtzzentrum f\"ur Schwerionenforschung GmbH, 64291 Darmstadt, Germany}
\author{R.\ Bollig}
\affiliation{Max-Planck-Institut f\"ur Astrophysik, Karl-Schwarzschild-Str.~1, 85748 Garching, Germany}
\affiliation{Physik Department, Technische Universit\"at M\"unchen, James-Franck-Str.~1, 85748 Garching, Germany}
\author{H.-T.\ Janka}
\affiliation{Max-Planck-Institut f\"ur Astrophysik, Karl-Schwarzschild-Str.~1, 85748 Garching, Germany}
\author{A.\ Schwenk}
\affiliation{Institut f\"ur Kernphysik, Technische Universit\"at Darmstadt, 64289 Darmstadt, Germany}
\affiliation{ExtreMe Matter Institute EMMI, GSI Helmholtzzentrum f\"ur Schwerionenforschung GmbH, 64291 Darmstadt, Germany}
\affiliation{Max-Planck-Institut f\"ur Kernphysik, Saupfercheckweg 1, 
69117 Heidelberg, Germany}

\begin{abstract}

Neutrino-pair production and annihilation through nucleon-nucleon
bremsstrahlung is included in current supernova simulations by rates
that are based on the one-pion-exchange approximation. Here we explore
the consequences of bremsstrahlung rates based on a modern nuclear
interactions for proto-neutron star cooling and the corresponding
neutrino emission. We find that despite a reduction of the
bremsstrahlung emission by a factor of 2--5 in the neutrinospheric
region, models with the improved treatment exhibit only $\lsim$5\%
changes of the neutrino luminosities and an increase of
$\lsim$0.7\,MeV of the average energies of the radiated neutrino
spectra, with the largest effects for the antineutrinos of all flavors
and at late times. Overall, the proto-neutron star cooling evolution
is slowed down modestly by $\lsim$0.5--1\,s.

\end{abstract}

\maketitle

\section{Introduction}
\label{sec:intro}

Neutrinos play an important role in core-collapse supernovae. Not only
will they allow us to probe the intererior of the next galactic
supernovae, they also carry away most of the energy liberated during
core collapse and deposit some of that energy in the region behind the
shock, thus possibly triggering the explosion (see,
e.g.,~\cite{Janka2007,Janka2012,Burrows2013,Foglizzo2015,Janka2016}
and references therein). A sound theoretical understanding of neutrino
interactions is therefore a key ingredient to realistic supernova
simulations.

Among the interactions involving nucleons, elastic scattering is the
main source of neutrino opacity. However, this does not change the
number of neutrinos nor their energy. The neutrino 
energy can be changed also by
inelastic scattering on interacting nucleons ($\nu NN \leftrightarrow
\nu NN $), while the closely related nucleon-nucleon ($NN$)
bremsstrahlung and its inverse, pair absorption, ($NN\leftrightarrow
NN\nu\bar\nu$) play an important role both for determining the cooling
of the newly formed neutron star and the neutrino spectra.

The bremsstrahlung rates used in supernova simulations are typically
based on analytical fit functions provided by Hannestad and
Raffelt~\cite{hannestad98} (HR) who treated the nucleon-nucleon
potential in the one-pion-exchange (OPE) approximation at Born level
and essentially considered the interactions among neutrons only.

Besides the bremsstrahlung rates, Hannestad and Raffelt
also provided an expression for inelastic scattering, however without
including recoil effects consistently in reactions on single and two
nucleons. Using the HR, Refs. \cite{Raffelt:2001kv,Keil:2002in} 
demonstrated that including energy transfers by inelastic scattering
($\nu NN \rightarrow \nu NN$) in addition to NN bremsstrahlung
and neutrino-pair absorption as well as energy transfers by nucleon
recoil ($\nu + N \rightarrow \nu + N$) has a negligible effect on
neutrino transport results in simulations.  For this reason inelastic
scattering was not taken into account in the development of VERTEX,
because this code considers the detailed nucleon recoil effects already.
Future work will have to investigate whether these conclusions also
hold for the case when inelastic scattering is included by more
sophisticated calculations and when recoil effects are included
consistently in reactions on single and two nucleons.

Recently, Bacca \textit{et al.}~\cite{bacca09,bacca12} studied $NN$
brems-strahlung using modern nuclear interactions based on chiral
effective field theory
(EFT)~\cite{weinberg90,weinberg91,Epel09RMP,Mach11PR}. This was
generalized to mixtures of neutrons and protons by Bartl \textit{et
al.}~\cite{bartl14}. In addition, they demonstrated the necessity to
go beyond the Born approximation at low densities and did so employing
a T-matrix-based formalism using as input phase shifts extracted from
experiment. For neutrons, bremsstahlung rates based on $NN$ scattering
were also developed previously by Hanhart \textit{et
al.}~\cite{Hanhart2001}.

In this paper, we investigate the influence of the bremsstrahlung rate
on the supernova and proto-neutron star evolution and the
corresponding neutrino emission by comparing results with our improved
treatment and the HR description, using one-dimensional simulations of
a 9.6\,$M_\odot$ and a 27\,$M_\odot$ progenitor, producing neutron
stars of about 1.25\,$M_\odot$ and 1.59\,$M_\odot$ (gravitational
mass), respectively.

After completion of our work we became aware of a similar study by
Fischer, however for an 18\,$M_\odot$ progenitor star giving birth to
a neutron star with a baryonic mass of 1.65\,$M_\odot$ (gravitational
mass of 1.45--1.5\,$M_\odot$)~\cite{Fischer2016}. In contrast to
our hydrodynamic evolution models, the hydrodynamic simulations by
Fischer do not include the effects of proto-neutron star convection.
Nevertheless, the main conclusions from both studies are basically in
agreement.

Our approach can be summarized as follows: We calculate
energy-averaged mean-free paths using the T-matrix-based formalism
from Ref.~\cite{bartl14} relative to one-pion-exchange results. We
use a parametrization of the temperature in terms of the density
and fixed values of the electron fraction to obtain a set of
one-dimensional fits for this ratio as a function of density.
We then implement this estimate of the improved rate in our
simulations by multiplying the HR rate by this ratio.

In Sec.~\ref{sec:brems} we briefly discuss the bremsstrahlung rate
used here and provide a simple analytical correction factor that
allows us to rescale the standard rate based on the simple OPE
ansatz. In Sec.~\ref{sec:setup} we describe the numerical setup of the
supernova simulations. We present our results in Sec.~\ref{sec:results}
mainly for simulations of one progenitor star. Finally, we conclude in
Sec.~\ref{sec:concl}.

\section{Structure factor for bremsstrahlung}
\label{sec:brems}

In Ref.~\cite{bartl14}, the formalism for bremsstrahlung rates in
mixtures of protons and neutrons was developed and a partial-wave
decomposition of the expression was done. We will work with this
expression and use $NN$ phase shifts extracted by the Nijmegen
partial-wave analysis~\cite{stoks93} in combination with the
T-matrix. At this level, the on-shell partial-wave-expanded matrix
elements are given in terms of the phase shifts $\delta_{lSJ}$ by
\begin{equation}
T_{lSJ}(\vec k,\vec k; E=\frac{k^2}{\mu})=-\frac{2\pi}{\mu} \frac{e^{2i\delta_{lSJ}}-1}{2ik} 
\end{equation}
in uncoupled and
\begin{align}
T_{ll'SJ}&(\vec k,\vec k; E=\frac{k^2}{\mu})=-\frac{2\pi}{\mu} \frac{1}{2ik}\notag\\&\times\begin{cases}
\left[e^{2i\delta_{lSJ}}\cos2\epsilon_J-1 \right] & \mbox{for }l=l' \,, \\
\left[ie^{i(\delta_{lSJ}+\delta_{l'SJ})}\sin2\epsilon_J-1 \right] & \mbox{for }l\neq l'
\end{cases}
\end{align}
in coupled channels. Here, $\vec k$ is the relative momentum and $\mu$
the reduced mass of the nucleons, $l$, $l'$ and $S$ are the orbital
angular momenta and the total spin of the nucleon pair, and
$\epsilon_J$ is the mixing angle for given total angular momentum $J$.
Note that in the case of OPE, the relaxation rates can be calculated
analytically by evaluating the spin traces in Ref.~\cite{bartl14}.

\subsection{Analytical correction factor}
\label{sec:ana}

For a first estimate of the impact that the findings in
Ref.~\cite{bartl14} have on the proto-neutron star cooling and the
corresponding neutrino emission, we calculate an analytical correction
factor
\begin{align}
r_{Y_e}(\rho)\equiv\frac{\left<\lambda^{-1}\right>(\rho,Y_e,T(\rho))}{\left<\lambda^{-1}\right>_{\rm OPEnn}(\rho,T(\rho))} \label{eq:correction_factor}
\end{align}
of the neutrino-antineutrino annihilation opacity relative to the
one-pion-exchange neutron-only (OPEnn) results, which are
conceptionally similar to Hannestad and Raffelt~\cite{hannestad98}. In
this first step, the correction factor is a function of density $\rho$
only. The temperature $T$ is parametrized by using
\begin{equation}
T(\rho) = T_\mathrm{SN}(\rho) \equiv
3 \mev \left(\frac{\rho}{10^{11}\gccm}\right)^{1/3} \,, \label{eq:Tdep}
\end{equation}
which was found to represent typical conditions in simulations
\cite{bacca12} (see also the discussion in Sec.~\ref{sec:results})
and the electron fraction $Y_e$ is treated as a parameter that only takes
fixed values (see Sec.~\ref{sec:setup}). The inverse mean-free path is
averaged over Boltzmann distributed neutrino and antineutrino
spectra. Both the T-matrix and OPEnn results are calculated using the
formalism discussed in Ref.~\cite{bartl14}, which assumes
nondegenerate conditions. Our results are then fitted by a function
of the form
\begin{equation}
r_{Y_e}(\rho)=a\ln(\rho)+10^{10}/\rho^{b}+c\,, 
\label{eq:fit}
\end{equation}
where $\rho$ is given in $\gccm$.

\begin{figure}[t]
\includegraphics[width=\columnwidth]{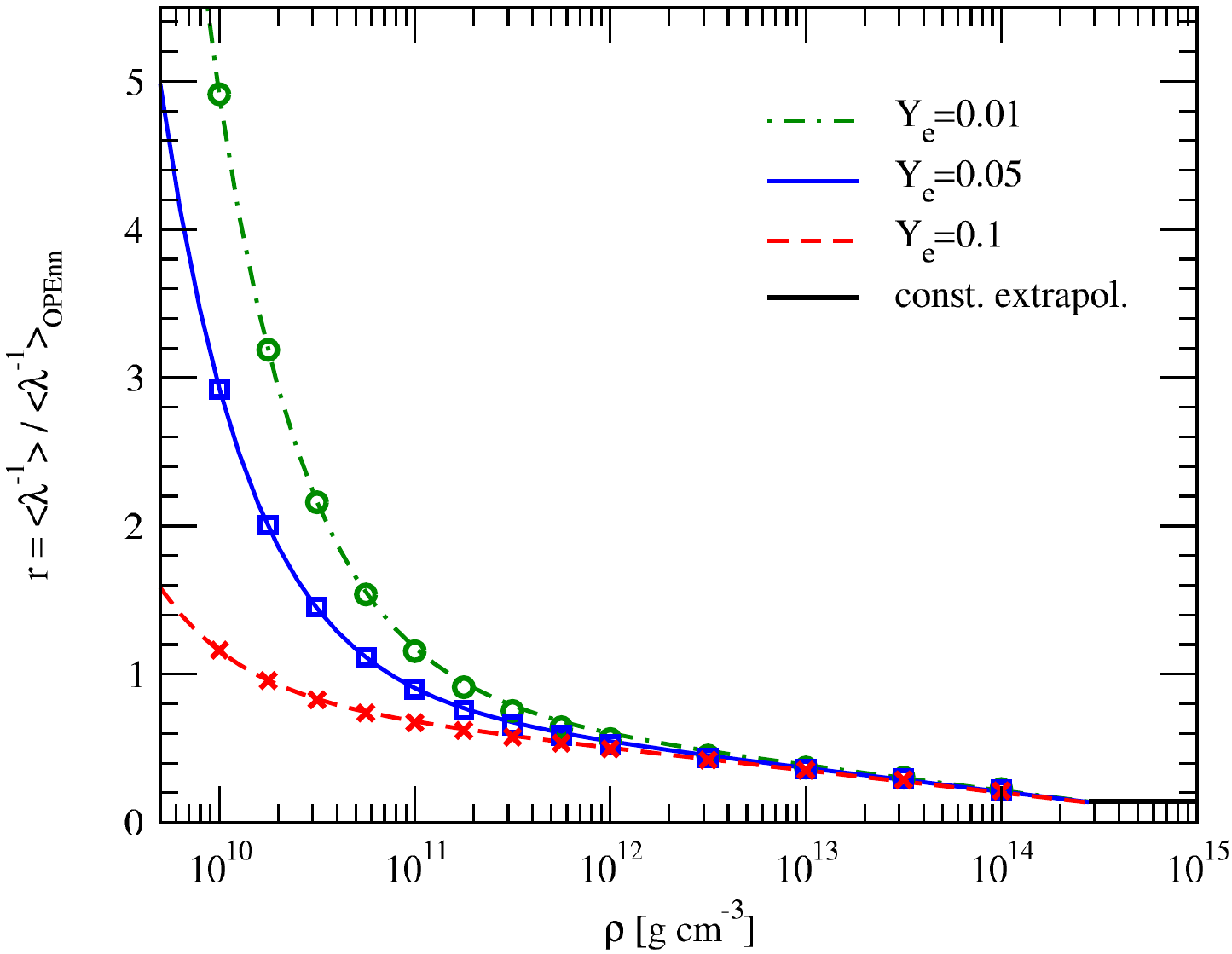}
\caption{Correction factor $r$ according to Eq.~\eqref{eq:correction_factor}
(symbols) and fit results (lines) for $Y_e=0.01, 0.05, \mbox{and }0.1$ as a 
function of density $\rho$. Also shown is the constant extrapolation beyond 
saturation density.}
\label{fig:fit}
\end{figure}

High-density rates are needed in the simulations, but beyond nuclear
saturation density, neutrinos are trapped and therefore reactions are
in equilibrium. In addition, our formalism breaks down at high
densities. As we do not expect $r$ to go to $0$, we extrapolate our
results with a constant, $r(\rho\geq\rho_0)=0.14$, to densities beyond
saturation density, $\rho_0 = 2.8\times10^{14}\gccm$. While the choice
is not well constrained, it is not expected to impact the simulation
due to the equilibrium conditions. For three relevant $Y_e$ values, we
show the fit parameters in Table~\ref{tab:fit} and a comparison of our
data points and fits in Fig.~\ref{fig:fit}.

\begin{table}[t]
\caption{Fit parameters for $r(\rho)$, Eq.~(\ref{eq:fit}), for $Y_e=0.01, 
0.05, \mbox{and }0.1$ shown in Fig.~\ref{fig:fit}.}
\begin{tabular}{c|ccc}
$Y_e$&$a$&$b$&$c$\\\hline
0.01 & $-$0.0649830 & 1.0446877 & 2.2954877\\
0.05 & $-$0.0685806 & 0.9680116 & 2.4176686\\
0.1  & $-$0.0726502 & 0.9395710 & 2.5558616
\end{tabular}
\label{tab:fit}
\end{table}

\subsection{Comparison with Hannestad and Raffelt}
\label{sec:hr}

As mentioned in Sec.~\ref{sec:ana}, we compute our T-matrix correction
factor relative to the OPE nn-only rate in our formalism, which is
conceptually similar to HR (obtained in the OPE approximation and
including protons as if they were neutrons). It does, however, exhibit
deviations from the actual HR result especially at high densities. One
possible explanation for this deviation is degeneracy. At
$10^{14}\gccm$, $T_\text{SN}/T_\text{F}\approx 1$, so degeneracy
effects start to contribute. Here, the Fermi temperature $T_\text{F}$
is the (neutron) Fermi energy in units of temperature. Our formalism
is purely non-degenerate (but was shown in pure neutron matter to be a
good approximation for partially-degenerate matter, see
Ref.~\cite{bacca12}), while the HR formalism interpolates between
degenerate and non-degenerate conditions.

\begin{figure}[t]
\includegraphics[width=\columnwidth]{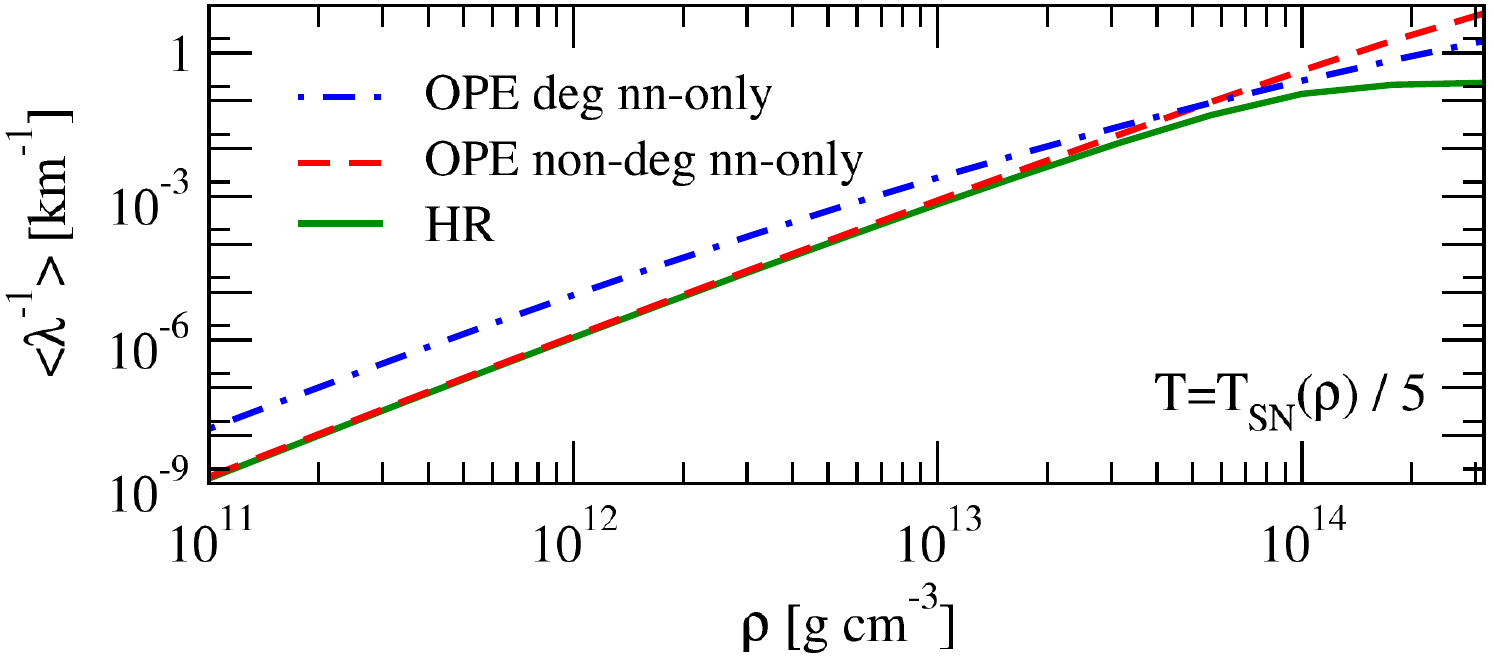}
\caption{Energy-averaged inverse mean-free path for pair annihilation 
by the bremsstrahlung process as a function of density, comparing HR
results~\cite{hannestad98} with our formalism in the degenerate and
non-degenerate limits. Note that the plot is based on a
temperature-density relation as given by Eq.~\eqref{eq:Tdep} but
reduced by a factor of 5.}
\label{fig:HRissue}
\end{figure}

In order to test this explanation, we make the conditions more
degenerate in Fig.~\ref{fig:HRissue} by reducing the temperature by a
factor of $5$ compared to the parametrization in
Eq.~\eqref{eq:Tdep}. At low densities, HR and the non-degenerate OPE
results agree very well. At $10^{14}\gccm$, $T/T_\text{F}\approx0.2$
and deviations between these are significant. A degenerate version of
our formalism is available~\cite{lykasov08,bacca09} which we would
expect to match the HR results at lower $T/T_\text{F}$. The opacity
obtained with this structure factor is also shown in
Fig.~\ref{fig:HRissue}. While it lies closer to the HR opacity at high
densities, there is still a significant deviation. The HR opacity
levels off and would eventually decrease if we increase the density
even further. This maximum can be moved to lower densities by making
the conditions even more degenerate. The handling of
multiple-scattering effects in the HR formalism seems to be the root
of this maximum, which we consider to be likely unphysical as a denser
medium should always be more opaque.

Using the comparison with OPE nn-only results has the advantage that
the assumptions and approximations of the formalism are the same and
our correction factor directly measures the impact of the more
advanced nuclear interactions compared to OPE. As our correction
factor is a crude approximation by construction, this strategy seems
justified.

\section{Numerical setup and simulated models}
\label{sec:setup}

Our simulations were performed with the one-dimensional version of the
\textsc{Prometheus-Vertex} code~\cite{Rampp:2002}, i.e., in spherical
symmetry. We use the most elaborate set of neutrino processes as
described in Ref.~\cite{Rampp:2002} with the improvements of
Ref.~\cite{Buras:2006} and further upgrades as reported in
Ref.~\cite{Mirizzi:2015eza}. In particular, our revised implementation
of charged-current neutrino-nucleon interactions (according to
Ref.~\cite{BurrowsSawyer:1998}) includes nucleon self-energy
corrections~\cite{MartinezPinedo:2012,Roberts:2012,Reddy:1998,Hempel:2015}.
We also account for differences in the weak-magnetism corrections of
neutral-current neutrino-nucleon scatterings as in
Ref.~\cite{Horowitz:2002} by a separate treatment of the transport of
$\nu_{\mu,\tau}$ and of $\bar\nu_{\mu,\tau}$.  Moreover, we take into
account proto-neutron star convection by a mixing-length treatment as
described in Ref.~\cite{Mirizzi:2015eza}.

The T-matrix modified bremsstrahlung rates are implemented using the
fit formula of Eq.~(\ref{eq:fit}) with the parameter values of
Table~\ref{tab:fit} to correct the HR rates employed in our standard
description. To handle the $Y_e$ dependence of the coarsely gridded
table data a step function in $Y_e$ space is used. Specifically, for
grid cells with $Y_e<0.05$ the tabulated fit for
$Y_{e,\text{table}}=0.01$ is applied, for grid cells with $0.05 \leq
Y_e<0.1$ the tabulated fit for $Y_{e,\text{table}}=0.05$, and finally
for grid cells with $Y_e \geq 0.1$ the tabulated fit for
$Y_{e,\text{table}} = 0.1$ is adopted.  This allows us to always test
the maximal influence of the T-matrix correction factor in the
low-density regime of $\rho<10^{11}$\,g\,cm$^{-3}$ as the strength of
the correction decreases with increasing $Y_e$, see
Fig.~\ref{fig:fit}. Conversely, for high-density proto-neutron star
conditions of $\rho>10^{12}$\,g\,cm$^{-3}$, where bremsstrahlung is
most relevant, the correction factors are nearly independent of $Y_e$
and the details of the handling of the tabulated data are less
important.  The chosen $Y_e$ correction factor is then applied to the
0th and 1st Legendre moments of the bremsstrahlung opacities in an
energy-independent way, because the correction factors are given as
ratios of the annihilation opacities for neutrino and antineutrino
pairs populating equilibrium phase-space distributions.
Since bremsstrahlung annihilation and $\nu\bar\nu$ pair
production are implemented by applying detailed-balance constraints,
neutrino equilibration is guaranteed to be numerically recovered.

We simulate the phases of stellar core-collapse, bounce, post-bounce
accretion and supernova explosion, and the subsequent cooling of the
proto-neutron star for up to more than 10\,s for a 9.6\,$M_\odot$
progenitor star (see
Refs.~\cite{Woosley:2015,MuellerJanka:2012,Melson:2015}) and a
27\,$M_\odot$ progenitor (see Ref.~\cite{Woosley:2002}). In the former
simulation, we employ the SFHo equation of state (EOS) for hot nuclear
matter of Ref.~\cite{SteinerHempel:2013}, in the latter case the LS220
EOS of Lattimer and Swesty~\cite{LS:1991} with incompressibility $K =
220$\,MeV.  Results of our simulations using the standard
implementation of $NN$ bremsstrahlung according to the HR
rates~\cite{hannestad98} were reported in Ref.~\cite{Mirizzi:2015eza}.

While the 9.6\,$M_\odot$ model as a low-mass progenitor with very
steep density gradient outside of the iron core explodes naturally
even in spherical symmetry~\cite{Melson:2015}, the explosion of the
27\,$M_\odot$ model is initiated artificially at 0.5\,s after core
bounce by reducing the pre-shock density and thus the
explosion-damping mass-accretion rate of the stalled shock gradually
by up to a factor of 30 (see also Ref.~\cite{Mirizzi:2015eza}).

\section{Results}
\label{sec:results}

\subsection{Post-processing opacities}
\label{sec:postproc}

Before discussing the impact of the correction factor on the
proto-neutron star cooling and neutrino emission, we evaluate the
quality of our approximation by post-processing the radial profiles
obtained in the simulation of the 27\,$M_\odot$ progenitor and
comparing the corrected HR results with the full T-matrix results.

\begin{figure*}[h!t]
\begin{center}
\includegraphics[width=.76\textwidth]{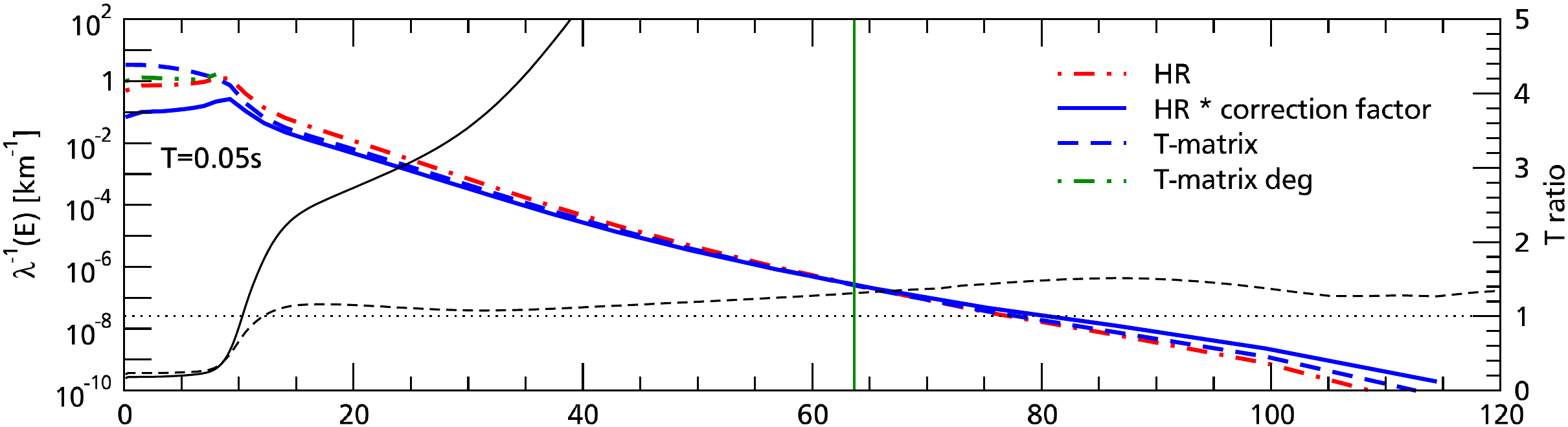}\\
\includegraphics[width=.76\textwidth]{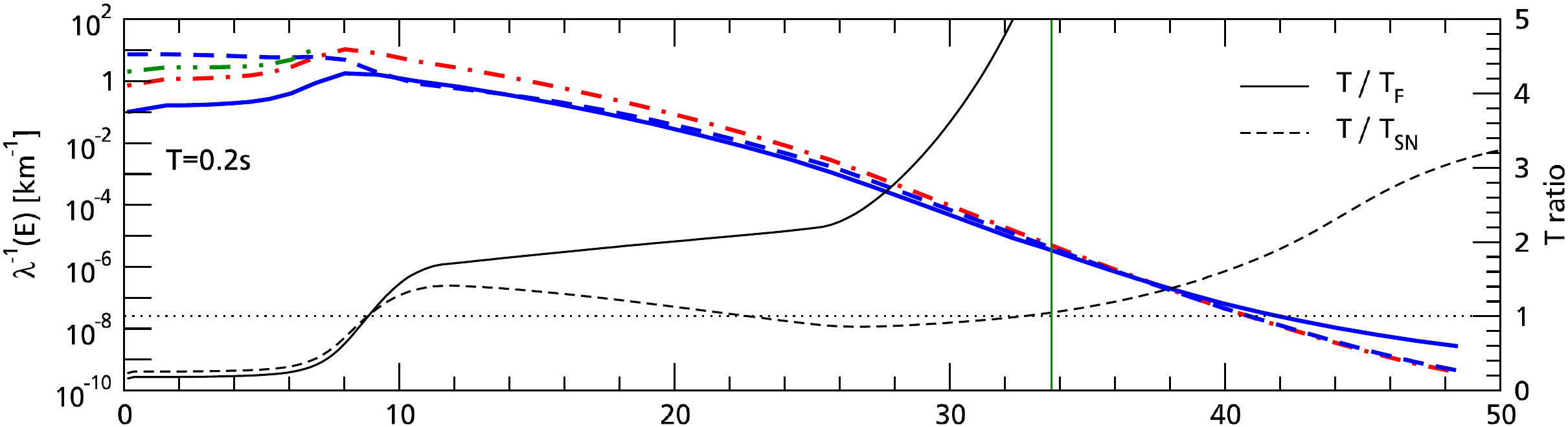}\\
\includegraphics[width=.76\textwidth]{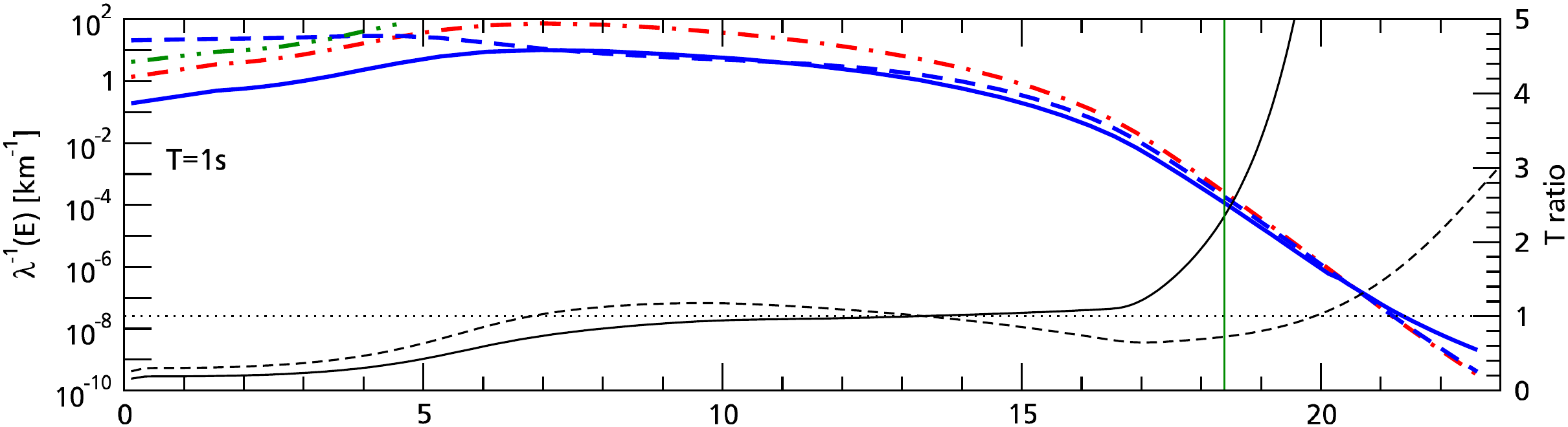}\\
\includegraphics[width=.76\textwidth]{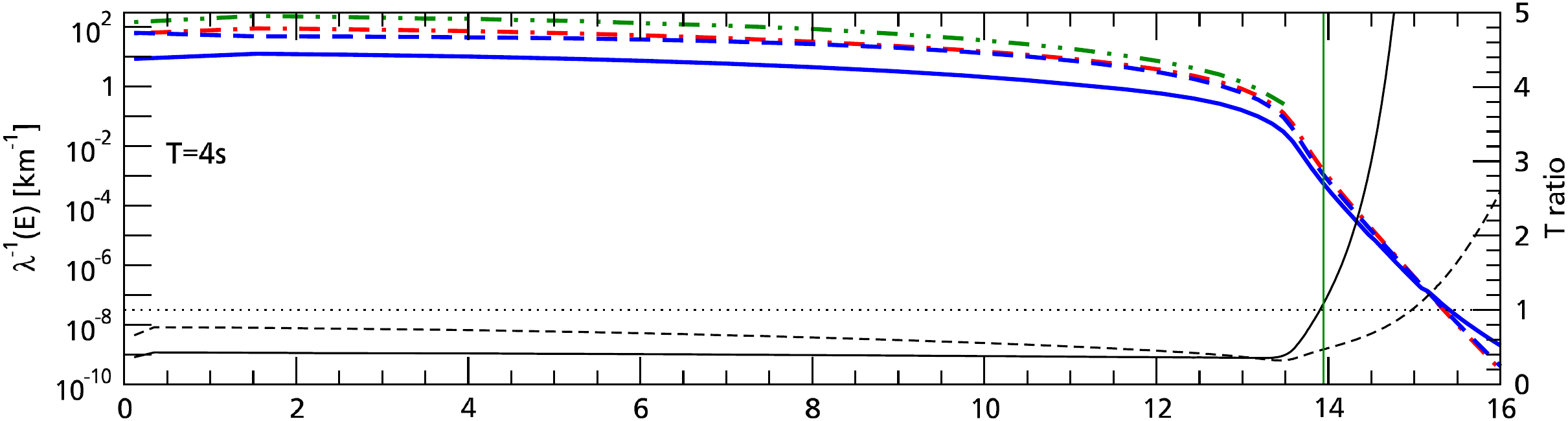}\\
\includegraphics[width=.76\textwidth]{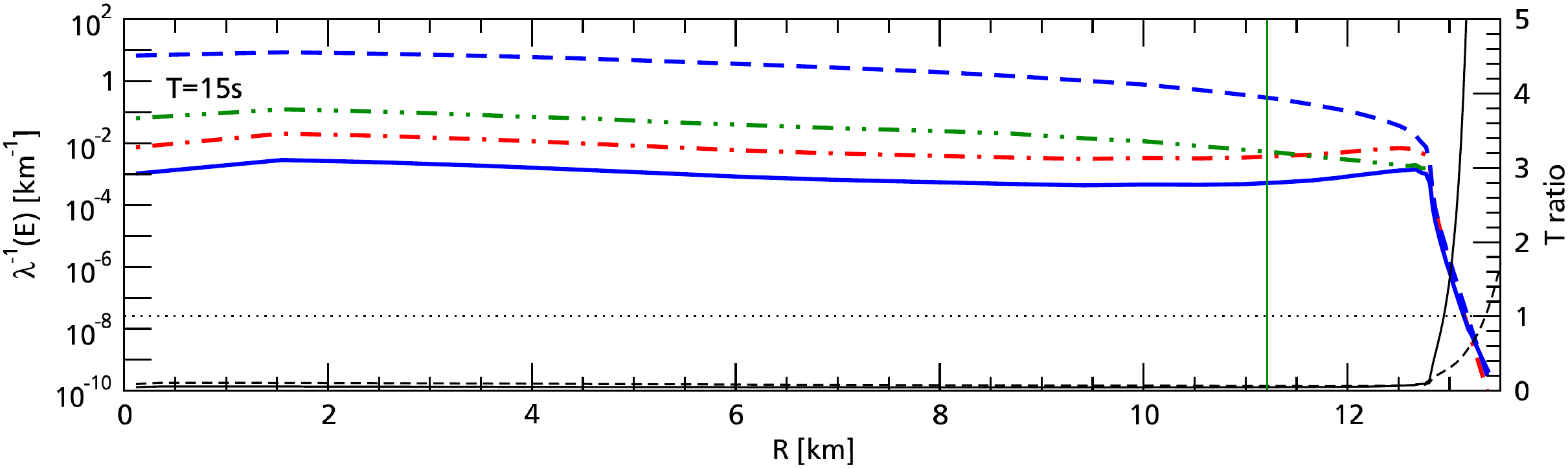}
\caption{Inverse mean-free path against pair absorption for a $\nu_x$ with 
local mean energy as a function of radius along profiles for our
27\,$M_\odot$ simulation at $50\,$ms, $200\,$ms, $1\,$s, $4\,$s, and
$15\,$s. Shown is the opacity used in the simulation (HR $\times$
correction factor, blue solid line) along with the HR (red dash-dotted
line) and T-matrix (blue dashed line) opacities. For comparison, we
also show degenerate T-matrix results calculated for pure neutron
matter. The black solid line shows the degeneracy $T/T_\mathrm{F}$,
the black dashed line shows $T/T_\mathrm{SN}$ [see
Eq.~\eqref{eq:Tdep}]. The green vertical line indicates the position
of the $\nu_x$ neutrinosphere.}
\label{fig:profiles}
\end{center}
\end{figure*}

\begin{figure}[t]
\includegraphics[width=\columnwidth]{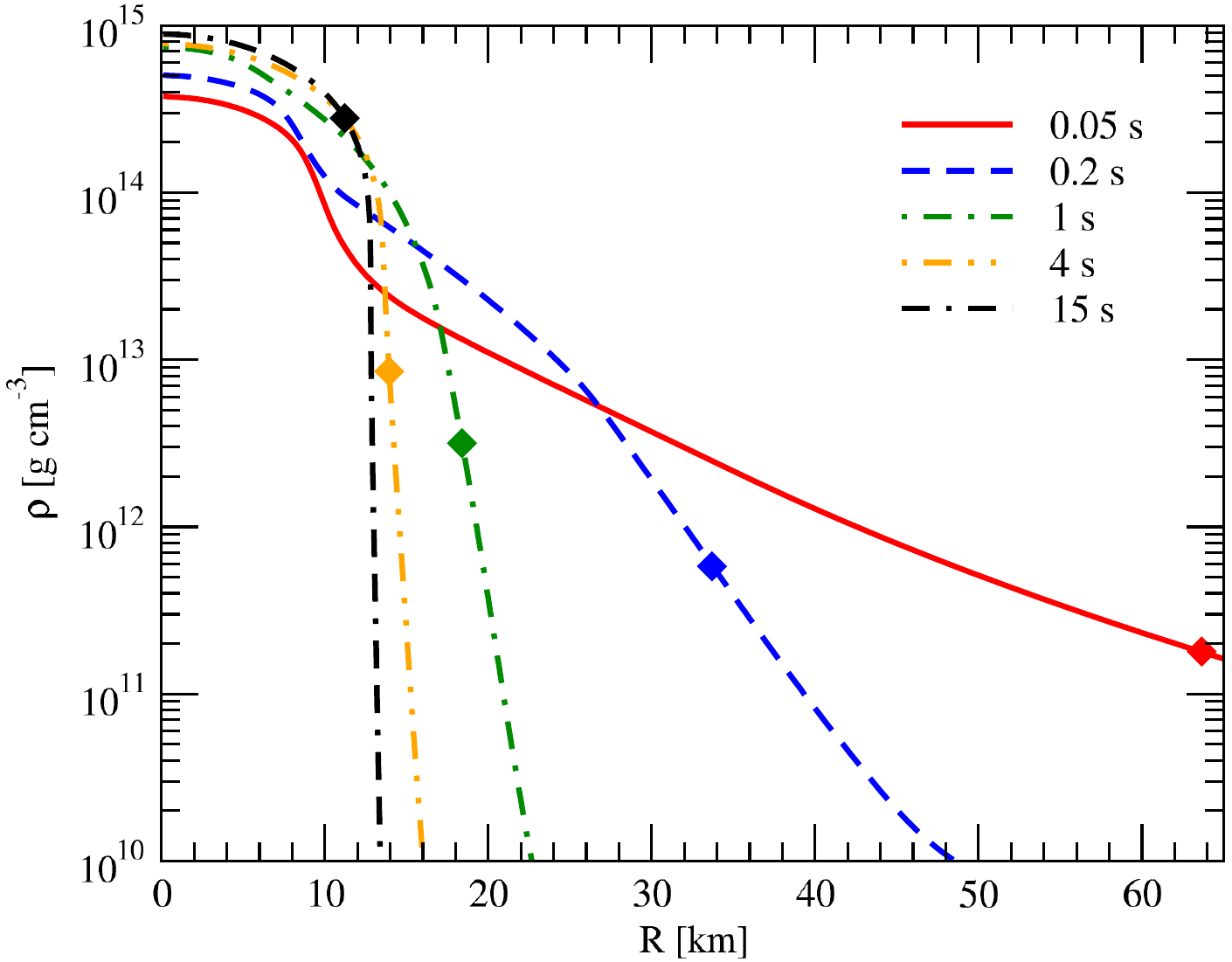}
\caption{Density versus radius of the proto-neutron star in our
27\,$M_\odot$ model at the times displayed in
Fig.~\ref{fig:profiles}. The colored diamonds mark the
position of the $\nu_x$ neutrinosphere (as also shown in
Fig.~\ref{fig:profiles}).}
\label{fig:densities}
\end{figure}

We do so in Fig.~\ref{fig:profiles}, where we show the inverse
mean-free path for a neutrino with $E=\left<E_{\nu_x}\right>$ against
pair annihilation, assuming a Fermi-Dirac distribution for the
antineutrino. Here, $\left<E_{\nu_x}\right>$ is the local mean
neutrino energy of muon/tau neutrinos as obtained in the
simulation. The plot range is determined by the radius where the
density drops below $10^{10}\gccm$. The $\nu_x$ neutrinosphere
position is indicated by the green vertical line. We define it by the
radius where the optical depth of a neutrino with local mean energy
becomes smaller than one, using the sum of the opacities of all kinds
of (in)elastic scattering and pair annihilation processes (averaged
for muon and tau neutrinos). This location roughly marks the region
where neutrinos decouple from the stellar medium, but it is neither
identical with the energy sphere nor with the transport
sphere~\cite{Janka:1995cu,Raffelt:2001kv}, which have to be introduced
for a detailed discussion of muon and tau neutrino transport. The
densities corresponding to the neutrinospheric positions according to
our (crude) definition can be extracted from Fig.~\ref{fig:densities},
where we present density profiles of the proto-neutron star
corresponding to the times picked for Fig.~\ref{fig:profiles}.

Ideally, the blue lines showing the approximated and full T-matrix
result should lie on top of each other. Looking at the earlier
profiles up to $1\,$s, we find good agreement between the two lines
over a wide density range and especially around the neutrinosphere
where the rates are most relevant.

We do find deviations, however, at small and large radii. The former
can be attributed to degeneracy effects, which are neither included in
our non-degenerate T-matrix formalism nor in the OPEnn calculations
used to fit the correction factor, but they are taken into account by
the HR rate. The black solid lines show the ratio of the temperature
over the Fermi temperature, $T/T_\mathrm{F}$, as an indicator of the
degeneracy. ($T_\mathrm{F}$ is the Fermi energy divided by the
Boltzmann constant.) Deviations appear where this value is
significantly below $1$. No formalism has been derived yet to
calculate bremsstrahlung rates in mixtures of neutrons and protons at
degenerate conditions using modern interactions. Nevertheless, we can
use the formalism developed in Refs.~\cite{lykasov08,bacca09} to
calculate T-matrix opacities in pure neutron matter under degenerate
conditions (in the region where $T/T_\mathrm{F}<1/\pi$). This can
explain some of the discrepancy, but the original HR result still lies
closer to the T-matrix opacities than the corrected one. This is
partly a result of the HR issues discussed in Sec.~\ref{sec:hr}.

The deviations found at small densities in the outer regions can be
attributed to temperature effects. Our fit factor is a one-dimensional
function of density, assuming temperature to be parametrized by
$T_\mathrm{SN}(\rho)$ given by Eq.~\eqref{eq:Tdep}. We plot
$T/T_\mathrm{SN}$ along the profiles in Figure~\ref{fig:profiles} and
see significant deviations from unity especially in the core and at
large radii.

At later times, the proto-neutron star has cooled and becomes highly
degenerate. In the outer regions, our approximation still works fine,
while in the center degeneracy effects lead to major deviations
between our non-degenerate T-matrix results and the corrected HR
opacities. As expected, this is significantly reduced when using the
degenerate T-matrix rate instead. However, the discrepancies remain
sizable.

Since neutrinos are in equilibrium in the high-density regions of the
proto-neutron star interior and free-streaming in the low-density
outer regions, the intermediate region around the neutrinosphere is
most important for both the proto-neutron star cooling and the
neutrino signal. In this region, our approximation works reasonably
well, except for very late times when the neutrinospheres lie inside
the degenerate proto-neutron star.

Furthermore, our corrected HR result tends to overestimate the
effects. Hence our first sensitivity study can be considered as a test
for the upper bounds on consequences of T-matrix modifications to the
bremsstrahlung process.

\begin{figure}[b]
\includegraphics[width=\columnwidth]{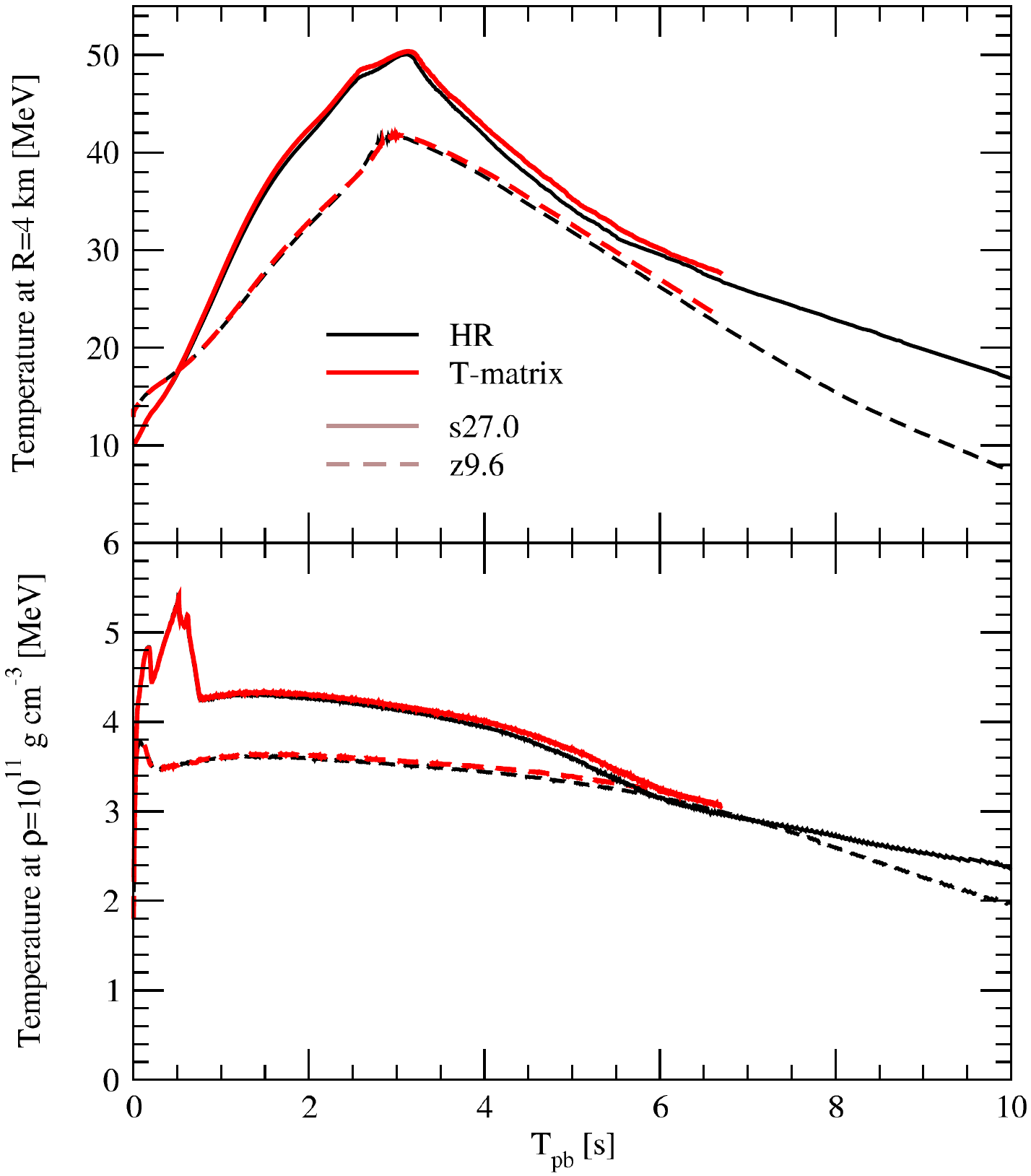}
\caption{Temperature in MeV in the innermost core of the nascent
neutron star (at a radius of 4\,km; top panel) and at a density of
10$^{11}$\,g\,cm$^{-3}$ (bottom panel) in our 27\,$M_\odot$ (solid
lines) and 9.6\,$M_\odot$ (dashed lines) models as functions of
time after bounce.}
\label{fig:temp}
\end{figure}

\begin{figure*}[h!t]
\includegraphics[width=.87\textwidth]{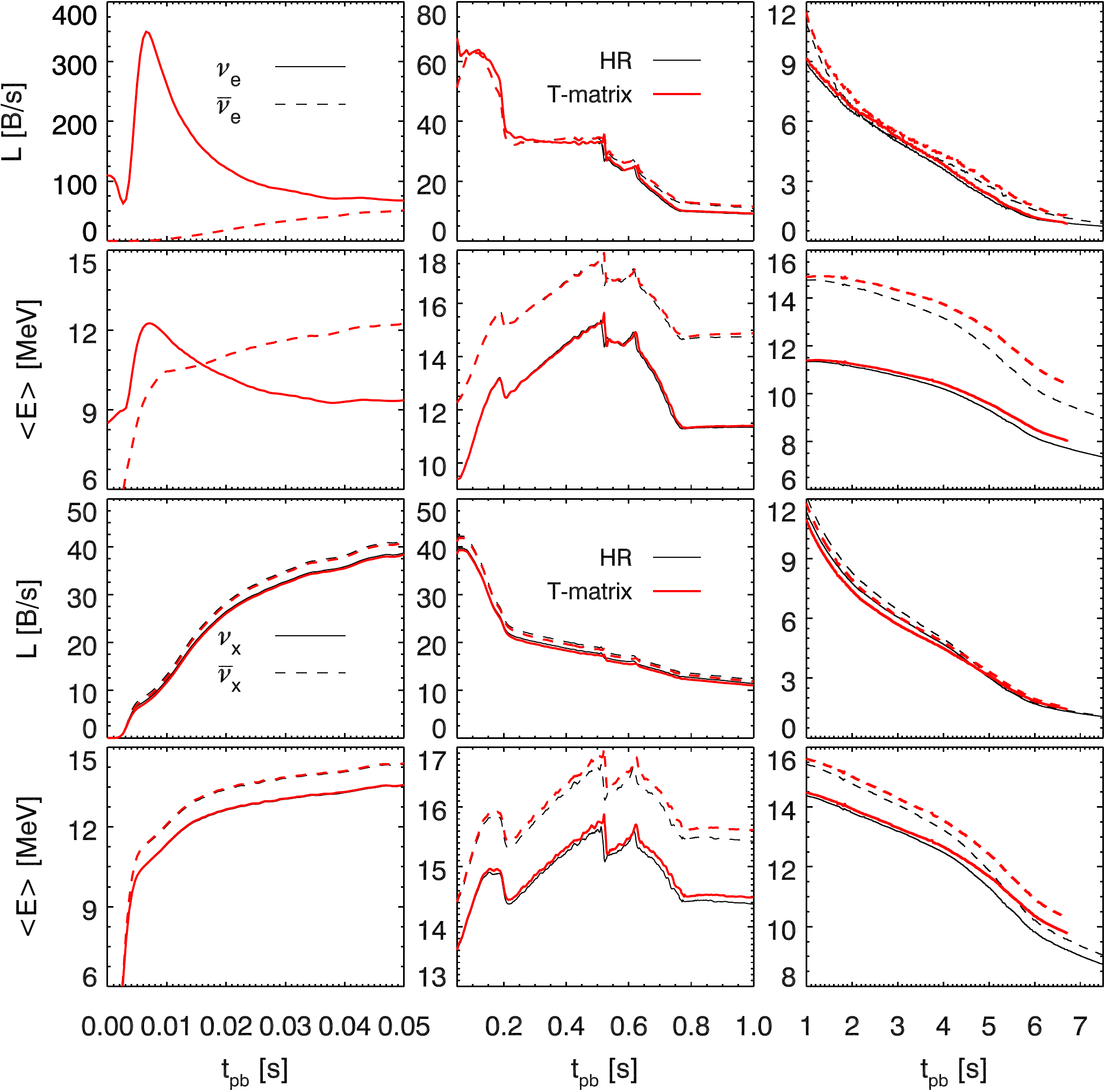}
\caption{Neutrino luminosities (in $10^{51}$\,erg\,s$^{-1}$; bethe/s or
B/s) and radiated mean neutrino energies for our 
27\,$M_\odot$ simulation as function of time after bounce,
evaluated at a radius of 500\,km for an observer in the lab frame 
at infinity. The left column shows the shock-breakout and $\nu_e$
burst phase, the middle column the accretion phase with the onset
of the supernova explosion, and the right column the neutrino signal
during the Kelvin-Helmholtz cooling phase of the newly formed neutron
star. The upper two rows provide the luminosities and mean energies
for $\nu_e$ and $\bar\nu_e$, the lower two rows those for heavy-lepton
neutrinos, $\nu_x$, and antineutrinos, $\bar\nu_x$.}
\label{fig:nusignal}
\end{figure*}

\subsection{Impact on proto-neutron star cooling and neutrino emission}

As mentioned in Sec.~\ref{sec:setup}, we have simulated the collapse
and explosion as well as the subsequent proto-neutron star cooling
phase for a 9.6\,$M_\odot$ and a 27\,$M_\odot$ progenitor.

The investigated 9.6\,$M_\odot$ progenitor can explode fairly easily
and rapidly by the neutrino-driven mechanism even in spherical
symmetry \cite{Melson:2015tia,Mirizzi:2015eza}, whereas explosions of
progenitors above $\sim$10\,$M_\odot$ (like the 27\,$M_\odot$ case
simulated here) require the support by multi-dimensional effects and
in spherically symmetric simulations (such as the ones performed in
this work) need to be triggered artificially.  In all cases, however,
including the 9.6\,$M_\odot$ star, multi-dimensional effects have a
strong influence on how and when the explosion develops. For this
reason it does not make any sense to discuss possible effects of a
modified treatment of the bremsstrahlung process on the explosion
mechanism on the basis of spherically symmetric simulations. We
therefore constrain our discussion here mostly on the differences
caused by the bremsstrahlung process during the proto-neutron star
cooling phase.

The most important consequence of the effective T-matrix rates [i.e.,
of the HR rates multiplied by the correction factor of
Eqs.~\eqref{eq:correction_factor} and \eqref{eq:fit}] is to reduce
the annihilation opacity of the bremsstrahlung process for neutrino
pairs inside of the nascent neutron star. This can be concluded from
Fig.~\ref{fig:fit}, where the correction factor drops below unity at
densities above $\sim$10$^{11}$\,g\,cm$^{-3}$, which is (roughly)
interior to the neutrinosphere of $\nu_e$. Correspondingly, also the
production rate of neutrino-antineutrino pairs through this reaction
is decreased, affecting mainly the emission of heavy-lepton neutrinos
($\nu_x$), which are not created by charged-current processes in the
absence of muons and tau leptons~\cite{Keil:2002in}. With the T-matrix
rates, we therefore expect a reduced emission of muon and tau
neutrinos and a corresponding delay of the cooling of the newly formed
neutron star.

This expectation is confirmed by Fig.~\ref{fig:temp}, which shows that
for both progenitor models the new-born neutron star becomes
slightly hotter in the innermost core but also cools more slowly, i.e.,
the temperature remains higher
for a longer time. It is important to note that this evolution
difference of the cooling proto-neutron star is initially triggered by
the reduced production of $\nu_x\bar\nu_x$ pairs via the effective
T-matrix rates in the neutrino-decoupling layers near the neutron-star
surface, but not by a change of the diffusion time scale of neutrinos
out of the dense interior of the neutron star. The diffusion time
scale is hardly affected by the modification of the bremsstrahlung
rate, because the total opacity is largely dominated by
neutral-current neutrino-nucleon scatterings and neutrino annihilation
by bremsstrahlung contributes only at a minor level.

\begin{figure*}[h!t]
\includegraphics[width=.87\textwidth]{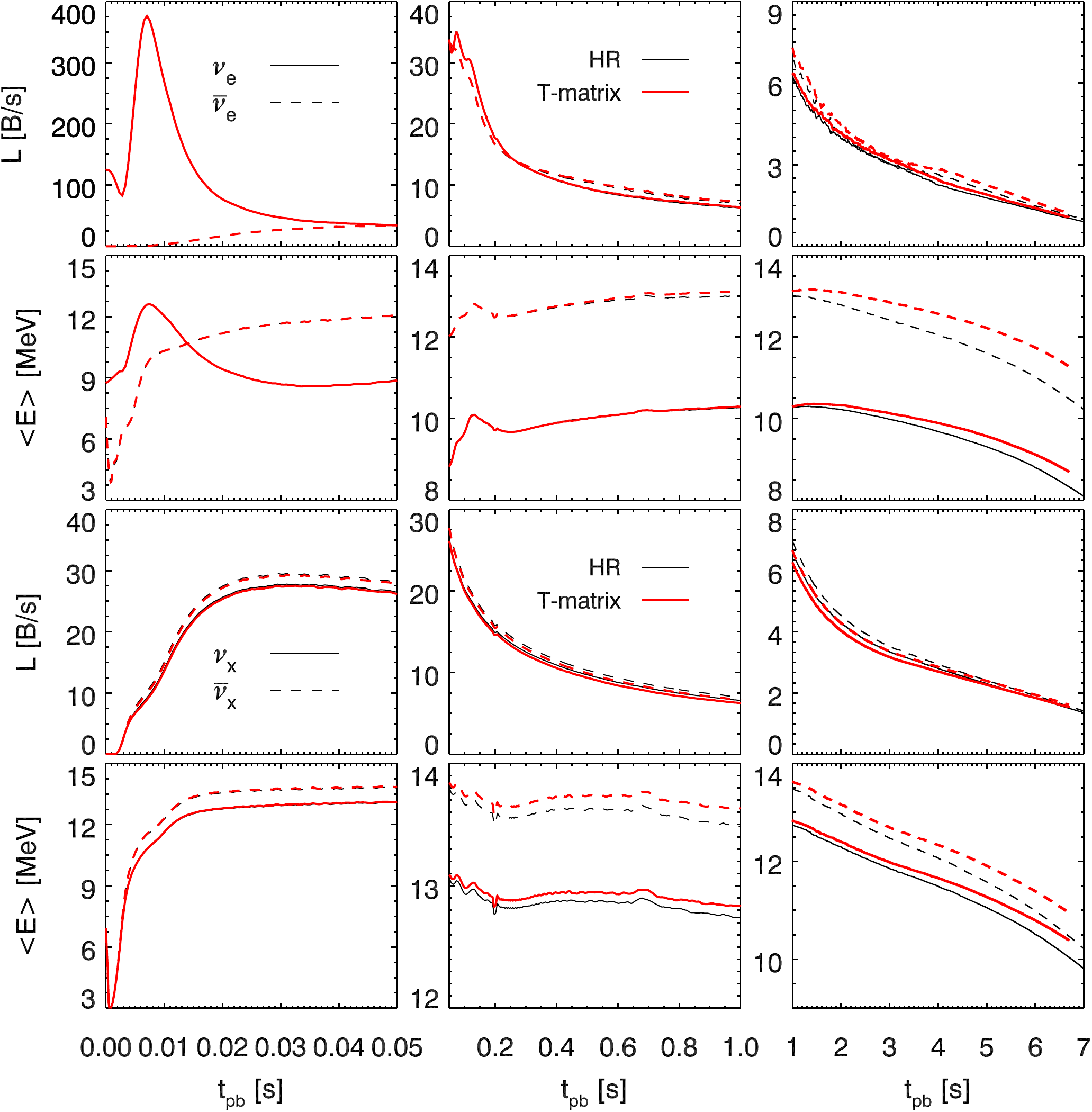}
\caption{Same as Fig.~\ref{fig:nusignal}, but for the $9.6\,M_\odot$
simulation}
\label{fig:nusignal_z9.6}
\end{figure*}

In agreement with this we observe very small differences of the
radiated neutrino luminosities and mean energies
(Figs.~\ref{fig:nusignal} and~\ref{fig:nusignal_z9.6})
during the accretion phase before the explosion is artificially
initiated at 0.5\,s after core bounce.
During this phase the neutrino emission is mainly produced in the hot
accretion mantle, where $e^\pm$ and $\nu_e\bar\nu_e$ annihilation
dominate the $\nu_x\bar\nu_x$ production. Only after accretion has
ended (when the explosion is well on its way, $t\gsim 0.8$\,s after
bounce), the $\nu_x$ and $\bar\nu_x$ emission is significantly
enhanced by the generation of these neutrinos through the
bremsstrahlung process, which dominates in the denser regions to which
the $\nu_x$ neutrinosphere retreats. Consequently, the differences of
the radiated neutrino luminosities and mean energies between the runs
with HR and T-matrix rates begin to grow gradually and become largest
at late times.

In detail, for the $27\,M_\odot$ progenitor we observe a slight reduction
of the luminosities of $\nu_{\mu,\tau}$ and $\bar\nu_{\mu,\tau}$ with
the T-matrix rates (Fig.~\ref{fig:nusignal}), while the radiated mean energies of these
neutrinos are larger by up to $\sim$0.5\,MeV with the biggest effects
at late times and slightly bigger for the antineutrinos
(Fig.~\ref{fig:nusignal}), which (due to the weak-magnetism
corrections) have lower nucleon-scattering opacities and decouple at a
deeper energy sphere.  These observations are compatible with the
decreased $\nu_x\bar\nu_x$ production by the T-matrix bremsstrahlung
rate, which moves the energy spheres of these neutrinos to higher
temperatures. On a much smaller level these effects can be observed
also during the accretion phase.

The $\nu_e$ and $\bar\nu_e$ emission properties exhibit hardly any
differences dependent on the bremsstrahlung treatment during the
accretion phase, where their emission originates from a
neutrinospheric region that is located within the hot accretion
layer. Only after accretion has ended ($t\gsim 0.8$\,s) differences
appear and grow gradually.  Figure~\ref{fig:nusignal} shows that the
T-matrix case, because of higher temperatures in the core of the
proto-neutron star as well as in its outer regions
(Fig.~\ref{fig:temp}), leads to an increase of the radiated $\nu_e$
and $\bar\nu_e$ luminosities by up to $\sim$4\% at $t \gsim 1$\,s
after bounce. Electron neutrinos and antineutrinos therefore take over
some of the energy transport from the flux-reduced heavy-lepton
neutrinos.  Simultaneously, the mean energies of $\nu_e$ and
$\bar\nu_e$ increase by up to $\sim$0.7\,MeV with the bigger
differences for $\bar\nu_e$ and at later times.  These differences
lead to an enhancement of the loss of $\nu_e$ number relative to the
loss of $\bar\nu_e$, accelerating the deleptonization of the nascent
neutron star.  This explains the higher inner-core temperatures seen
in Fig.~\ref{fig:temp} by resistive heating (converting degeneracy
energy of electrons to thermal energy by
down-scattering~\cite{Burrows:1986me}). At the same time the total
neutrino luminosity (i.e., the sum of the luminosities of all neutrino
species) is decreased and the proto-neutron star cooling takes
correspondingly longer. The growing differences of the neutrino
emission at late times therefore are not directly caused by the
instantaneous differences of the bremsstrahlung rates of the HR and
T-matrix calculations. Instead, they mainly reflect the differences of
the neutron-star temperature and lepton-number profiles, which diverge
more and more as time goes on and as the time-integrated effects of
different bremsstrahlung treatments accumulate.

While we have displayed and discussed the results of our simulations
for the 27\,$M_\odot$ progenitor, which gives birth to a neutron star
with $\sim$1.776 ($\sim$1.592)\,$M_\odot$ baryonic (final
gravitational) mass~\cite{Mirizzi:2015eza}, the differences in the
bremsstrahlung rates were found to have very similar effects in the
case of the 9.6\,$M_\odot$ star, whose explosion leaves behind a
neutron star with a baryonic (gravitational) mass of $\sim$1.363
(1.252)\,$M_\odot$, see Fig.~\ref{fig:nusignal_z9.6}.  Overall,
the luminosity decline of $\nu_e$ and $\bar\nu_e$ is delayed with
the T-matrix rates by at most a few 100\,ms, whereas the mean
energies of $\bar\nu_e$ decrease to the same cooling level at late
times only with a delay of up to $\sim$1\,s.

\section{Conclusions}
\label{sec:concl}

We have explored the impact of new T-matrix results for $NN$
bremsstrahlung in mixtures of neutrons and protons from
Ref.~\cite{bartl14} in simulations of core-collapse
supernovae. Comparing the T-matrix results to the standard HR
rate~\cite{hannestad98}, we have developed a simple correction factor
as a function of density that can be used in simulations. This allowed
for a first estimate of how bremsstrahlung rates with modern nuclear
interactions have an impact on the cooling evolution and neutrino
emission of newly formed, hot neutron stars. Our approximation is
constructed such that it tends to produce upper bounds on the possible
changes compared to the case using the HR bremsstrahlung rate.

Because of the lack of charged-current production by beta-reactions
with nucleons in the absence of muons and tauons,
neutrino-antineutrino pair production and annihilation by the
bremsstrahlung process affects mostly the transport of heavy-lepton
neutrinos, $\nu_x$, in the supernova core and in newly formed neutron
stars. Any corresponding effect on the transport and emission of
$\nu_e$ and $\bar\nu_e$ is only indirect through changes of the
dynamical and thermal evolution of the nascent neutron star as a
consequence of alterations of the $\nu_x$ transport. An exception to
this fact may be the emission of $\bar\nu_e$ at very late times, when
the medium of the proto-neutron star becomes progressively
neutron-dominated and degenerate, in which case the production of
$\bar\nu_e$ by beta-reactions diminishes and bremsstrahlung becomes
relatively more important also for $\bar\nu_e$.

The differences in the neutrino emission for our modified
bremsstrahlung treatment compared to the HR rate show up mainly after
the post-bounce accretion has ended due to the onset of the supernova
explosion, i.e., during the Kelvin-Helmholtz cooling of the new-born
neutron star.  Our calculations imply a reduction of the
bremsstrahlung rate to $\sim$60--20\% of the reference case based on
the HR rate in the density regime between
$\sim$10$^{12}$\,g\,cm$^{-3}$ and $\sim$10$^{14}$\,g\,cm$^{-3}$
(Fig.~\ref{fig:fit}), where the $\nu_x$ neutrinospheres are located
during most of the cooling evolution of the compact remnant
(Fig.~\ref{fig:densities}).

With this reduction we find a modest stretching of the
neutrino-cooling time (by $\sim$0.5--1\,s) of the nascent neutron
star. Because of the reduced production of heavy-lepton neutrinos, the
$\nu_x$ luminosities decrease by up to $\sim$5\%, whereas higher
temperatures in the neutron star lead to an increase of the $\nu_e$
and $\bar\nu_e$ luminosities also by a few percent. The hotter neutron
star emits all neutrino species with slightly harder spectra. The
biggest effect can be seen for $\bar\nu_e$ and $\bar\nu_x$, whose mean
energies of the radiated spectra are higher by up to more than
0.5\,MeV at late times.  The differences of the neutrino-emission
properties between both bremsstrahlung treatments grow with time. This
is mainly a consequence of the accumulating differences in the thermal
structure of the neutron stars due to their divergent evolution, and
it is less caused by larger instantaneous differences of the
bremsstrahlung rates in the late (more degenerate) stages.

Despite the slightly higher mean energies of the radiated $\bar\nu_e$,
the neutrino-driven baryonic wind of the nascent neutron stars
exhibits an insignificantly lower (by $\lsim$0.01) electron (proton)
fraction and remains proton-rich at all times as reported in
Ref.~\cite{Mirizzi:2015eza}, which disfavors r-process nucleosynthesis
in the wind ejecta.  A similarly weak impact of the different
treatments of $NN$ bremsstrahlung on the characteristics of the
neutrino-driven wind was also reported from the independent study by
Fischer~\cite{Fischer2016}, who observed a reduction of the wind-$Y_e$
by at most $\sim$0.004. This is compatible with our findings, although
the time evolution of this outflow property differs considerably from
the one obtained in our simulations. Because of the omission of
convection inside the newly formed neutron star, the wind ejecta in
Fischer's models evolve from slightly proton-rich to slightly
neutron-rich conditions within a fraction of a second to dive through
a flat minimum between 2 and 3 seconds and to continuously increase
subsequently.  In contrast, including convection in our proto-neutron
star cooling calculations accelerates the neutronization of the
compact remnant. Therefore the wind-$Y_e$ in our models is on the
proton-rich side all the time and evolves through a broad hump of
several seconds duration, corresponding to the main period of
deleptonization, before it joins the late-time trend of a monotonic
rise seen in Fischer's simulations (the reader is referred to the
discussion of this behavior in Ref.~\cite{Mirizzi:2015eza}).

One may wonder why a reduction of the bremsstrahlung rates by factors
of 2--5 in the relevant region around the neutrinospheres does not
have a stronger impact on the cooling history and neutrino
emission. The reason for this modest reaction are compensating
effects, which are very typical of the considered multi-component
system with its tightly and nonlinearly coupled ingredients, whose
response to variations can damp and balance consequences of changes of
individual ingredients. In the considered case, the increasing
temperatures in the neutron star for reduced bremsstrahlung lead to
more $\nu_x\bar\nu_x$ creation by $e^\pm$ and $\nu_e\bar\nu_e$ pair
annihilation (with an energy production rate rocketing with
$\sim$$T^9$) so that these latter processes nearly completely
compensate the decrease of the emission by the bremsstrahlung
process. We also point out that changes of the bremsstrahlung rate
below a density of $\sim$10$^{11}$\,g\,cm$^{-3}$ have hardly any
influence, because at such low densities neutrinos begin to possess
very large mean free paths and therefore make the transition to free
streaming. On the other side, at densities above nuclear saturation
density modifications of the bremsstrahlung rate also have little
influence because at such densities neutrinos are in chemical
equilibrium nearly until the end of the optically thick
neutrino-cooling evolution. The equilibration is achieved also by
other processes like $e^\pm$ annihilation and the plasmon-neutrino
process, and the exact time scale to establish this equilibrium is not
relevant as long as it is much shorter than the evolution time scales
for contraction, cooling, and deleptonization of the nascent neutron
star. Since neutral-current neutrino-nucleon scattering dominates the
total neutrino opacity by far, the minor contributions from
bremsstrahlung annihilation (and possible changes) also have hardly
an effect on the diffusion time scale of neutrinos out of the neutron star.

In Ref.~\cite{bartl14}, calculations based on chiral EFT interactions
at next-to-next-to-next-to-leading order were shown to produce results
very similar to the T-matrix rates at the densities found to be
relevant in this paper. Hence, we expect our conclusions to hold for
chiral EFT interactions as well. This is of particular interest, as
chiral EFT interactions can be used to calculate and constrain the EOS
(for recent work, see
Refs.~\cite{hebeler13,krueger13,Holt2013,Carbone2013,Wellenhofer2015,drischler16}),
eventually allowing for a consistent treatment of neutrino
interactions and the EOS.

For a more consistent treatment of bremsstrahlung than in these
explorative supernova and proto-neutron star cooling simulations, the
temperature dependence of the rates has to be taken into account
explicitly. In this case, a simple correction factor like the one used
in our work cannot be defined as easily, but tabulated structure
factors may be preferable. In addition, an interpolation is required
between our non-degenerate and degenerate formalisms, and the latter
needs to be extended to mixtures of neutrons protons.

\begin{acknowledgments}

We thank Almudena Arcones, Lorenz H\"udepohl, Andreas Marek, Hannah Yasin,
and especially Bernhard M\"uller, who contributed to the original implementations
of nucleon self-energy corrections and of the mixing-length treatment of
proto-neutron star convection in \textsc{Prometheus-Vertex}. This work
was supported by the Deutsche Forschungsgemeinschaft through
Grant SFB 1245 and the Cluster of Excellence EXC~153 ``Origin and Structure
of the Universe’' (http://www.universe-cluster.de), the European Research Council
AdG No.~341157-COCO2CASA and Grant No.~307986 STRONGINT, as
well as the Studienstiftung des Deutschen Volkes. The numerical simulations were
carried out on the IBM iDataPlex system {\em hydra} of the Max Planck
Computing and Data Facility (MPCDF).
We also thank the Department of Energy's Institute for Nuclear
Theory at the University of Washington for its hospitality and
the Department of Energy for partial support during the
completion of this work.

\end{acknowledgments}


\begin{thebibliography}{10}

\bibitem{Janka2007}
H.-T. {Janka}, K.~{Langanke}, A.~{Marek}, G.~{Mart{\'{\i}}nez-Pinedo}, and
  B.~{M{\"u}ller},
\newblock Phys. Rept. {\bf 442}, 38 (2007).

\bibitem{Janka2012}
H.-T. {Janka},
\newblock Annu. Rev. Nucl. Part. Sci. {\bf 62}, 407 (2012).

\bibitem{Burrows2013}
A.~{Burrows},
\newblock Rev. Mod. Phys. {\bf 85}, 245 (2013).

\bibitem{Foglizzo2015}
T.~{Foglizzo} {\em et~al.},
\newblock Publications of the Astronomical Society of Australia {\bf 32}, e009
  (2015).

\bibitem{Janka2016}
H.-T. {Janka}, T.~{Melson}, and A.~{Summa},
\newblock arxiv:1602.05576.

\bibitem{hannestad98}
S.~Hannestad and G.~Raffelt,
\newblock Astrophys. J. {\bf 507}, 339 (1998).

\bibitem{Raffelt:2001kv}
G.~G. Raffelt,
\newblock Astrophys. J. {\bf 561}, 890 (2001).
%%CITATION = ASTRO-PH/0105250;%%

\bibitem{Keil:2002in}
M.~T. Keil, G.~G. Raffelt, and H.-T. Janka,
\newblock Astrophys. J. {\bf 590}, 971 (2003).
%%CITATION = ASTRO-PH/0208035;%%

\bibitem{bacca09}
S.~Bacca, K.~Hally, C.~J. Pethick, and A.~Schwenk,
\newblock Phys. Rev. C {\bf 80}, 032802 (2009).

\bibitem{bacca12}
S.~Bacca, K.~Hally, M.~Liebendörfer, A.~Perego, C.~J. Pethick, and A.~Schwenk,
\newblock Astrophys. J. {\bf 758}, 34 (2012).

\bibitem{weinberg90}
S.~Weinberg,
\newblock Phys. Lett. B {\bf 251}, 288  (1990).

\bibitem{weinberg91}
S.~Weinberg,
\newblock Nucl. Phys. B {\bf 363}, 3  (1991).

\bibitem{Epel09RMP}
E.~Epelbaum, H.-W. Hammer, and U.-G. Mei{\ss}ner,
\newblock Rev. Mod. Phys. {\bf 81}, 1773 (2009).
%%CITATION = ARXIV:0811.1338;%%

\bibitem{Mach11PR}
R.~Machleidt and D.~R. Entem,
\newblock Phys. Rept. {\bf 503}, 1 (2011).
%%CITATION = ARXIV:1105.2919;%%

\bibitem{bartl14}
A.~Bartl, C.~J. Pethick, and A.~Schwenk,
\newblock Phys. Rev. Lett. {\bf 113}, 081101 (2014).
%%CITATION = ARXIV:1403.4114;%%

\bibitem{Hanhart2001}
C.~Hanhart, D.~R. Phillips, and S.~Reddy,
\newblock Phys. Lett. B {\bf 499}, 9 (2001).
%%CITATION = ASTRO-PH/0003445;%%

\bibitem{Fischer2016}
T.~Fischer,
\newblock {arXiv:1608.05004}.

\bibitem{stoks93}
V.~G.~J. Stoks, R.~A.~M. Klomp, M.~C.~M. Rentmeester, and J.~J. de~Swart,
\newblock Phys. Rev. C {\bf 48}, 792 (1993),
\newblock \mbox{\url{http://nn-online.org}}.

\bibitem{lykasov08}
G.~I. Lykasov, C.~J. Pethick, and A.~Schwenk,
\newblock Phys. Rev. C {\bf 78}, 045803 (2008).

\bibitem{Rampp:2002}
M.~{Rampp} and H.-T. {Janka},
\newblock Astron.~Astrophys. {\bf 396}, 361 (2002).

\bibitem{Buras:2006}
R.~{Buras}, M.~{Rampp}, H.-T. {Janka}, and K.~{Kifonidis},
\newblock Astron.~Astrophys. {\bf 447}, 1049 (2006).

\bibitem{Mirizzi:2015eza}
A.~Mirizzi, I.~Tamborra, H.-T. Janka, N.~Saviano, K.~Scholberg, R.~Bollig,
  L.~H{\"u}depohl, and S.~Chakraborty,
\newblock Riv. Nuovo Cim. {\bf 39}, 1 (2016).
%%CITATION = ARXIV:1508.00785;%%

\bibitem{BurrowsSawyer:1998}
A.~Burrows and R.~F. Sawyer,
\newblock Phys. Rev. C {\bf 58}, 554 (1998).

\bibitem{MartinezPinedo:2012}
G.~Mart\'{\i}nez-Pinedo, T.~Fischer, A.~Lohs, and L.~Huther,
\newblock Phys. Rev. Lett. {\bf 109}, 251104 (2012).

\bibitem{Roberts:2012}
L.~F. Roberts, S.~Reddy, and G.~Shen,
\newblock Phys. Rev. C {\bf 86}, 065803 (2012).

\bibitem{Reddy:1998}
S.~Reddy, M.~Prakash, and J.~M. Lattimer,
\newblock Phys. Rev. D {\bf 58}, 013009 (1998).

\bibitem{Hempel:2015}
M.~Hempel,
\newblock Phys. Rev. C {\bf 91}, 055807 (2015).

\bibitem{Horowitz:2002}
C.~J. Horowitz,
\newblock Phys. Rev. D {\bf 65}, 043001 (2002).

\bibitem{Woosley:2015}
S.~E. Woosley and A.~Heger,
\newblock Astrophys. J. {\bf 810}, 34 (2015).

\bibitem{MuellerJanka:2012}
B.~Müller, H.-T. Janka, and A.~Heger,
\newblock Astrophys. J. {\bf 761}, 72 (2012).

\bibitem{Melson:2015}
T.~Melson, H.-T. Janka, and A.~Marek,
\newblock Astrophys. J. Lett. {\bf 801}, L24 (2015).

\bibitem{Woosley:2002}
S.~E. {Woosley}, A.~{Heger}, and T.~A. {Weaver},
\newblock Rev. Mod. Phys. {\bf 74}, 1015 (2002).

\bibitem{SteinerHempel:2013}
A.~W. Steiner, M.~Hempel, and T.~Fischer,
\newblock Astrophys. J. {\bf 774}, 17 (2013).

\bibitem{LS:1991}
J.~M. {Lattimer} and F.~D. {Swesty},
\newblock Nucl. Phys. A {\bf 535}, 331 (1991).

\bibitem{Janka:1995cu}
H.~T. Janka,
\newblock Astropart. Phys. {\bf 3}, 377 (1995).
%%CITATION = ASTRO-PH/9503068;%%

\bibitem{Melson:2015tia}
T.~Melson, H.-T. Janka, and A.~Marek,
\newblock Astrophys. J. {\bf 801}, L24 (2015).
%%CITATION = ARXIV:1501.01961;%%

\bibitem{Burrows:1986me}
A.~Burrows and J.~M. Lattimer,
\newblock Astrophys. J. {\bf 307}, 178 (1986).
%%CITATION = ASJOA,307,178;%%

\bibitem{hebeler13}
K.~Hebeler, J.~M. Lattimer, C.~J. Pethick, and A.~Schwenk,
\newblock Astrophys. J. {\bf 773}, 11 (2013).
%%CITATION = ARXIV:1303.4662;%%

\bibitem{krueger13}
T.~Krüger, I.~Tews, K.~Hebeler, and A.~Schwenk,
\newblock Phys. Rev. C {\bf 88}, 025802 (2013).
%%CITATION = ARXIV:1304.2212;%%

\bibitem{Holt2013}
J.~W. Holt, N.~Kaiser, and W.~Weise,
\newblock Prog. Part. Nucl. Phys. {\bf 73}, 35 (2013).
%%CITATION = ARXIV:1304.6350;%%

\bibitem{Carbone2013}
A.~Carbone, A.~Rios, and A.~Polls,
\newblock Phys. Rev. C {\bf 88}, 044302 (2013).
%%CITATION = ARXIV:1307.1889;%%

\bibitem{Wellenhofer2015}
C.~Wellenhofer, J.~W. Holt, and N.~Kaiser,
\newblock Phys. Rev. C {\bf 92} (2015).
%%CITATION = ARXIV:1504.00177;%%

\bibitem{drischler16}
C.~Drischler, K.~Hebeler, and A.~Schwenk,
\newblock Phys. Rev. C {\bf 93}, 054314 (2016).
%%CITATION = ARXIV:1510.06728;%%

\end{thebibliography}
\end{document}